\documentclass[review]{elsarticle}

\usepackage{lineno,hyperref}
\modulolinenumbers[5]

\journal{Journal of \LaTeX\ Templates}

\begin{document}

\begin{frontmatter}

\title{Processing optimization with parallel computing for 
the J-PET tomography scanner}

\author[SWIERK0]{W.~Krzemie\'n}
\author[WFAIS]{M.~Bala}
\author[WFAIS]{T.~Bednarski}
\author[WFAIS]{P.~Bia\l as}
\author[WFAIS]{E.~Czerwi\'nski}
\author[WFAIS]{A.~Gajos}
\author[UMCS]{M.~Gorgol}
\author[UMCS]{B.~Jasi\'nska}
\author[WFAIS]{D.~Kami\'nska}
\author[WFAIS,PAN]{\L .~Kap\l on}
\author[WFAIS]{G.~Korcyl} 
\author[SWIERK]{P.~Kowalski}
\author[WFAIS]{T.~Kozik}
\author[WFAIS]{E.~Kubicz}
\author[WFAIS]{Sz.~Nied\'zwiecki}
\author[WFAIS]{M.~Pa\l ka}
\author[SWIERK]{L.~Raczy\'nski}
\author[WFAIS]{Z.~Rudy}
\author[WFAIS]{O.~Rundel}
\author[WFAIS]{N.G.~Sharma}
\author[WFAIS]{M.~Silarski}
\author[WFAIS]{A.~S\l omski} 
\author[WFAIS]{K.~Stola} 
\author[WFAIS]{A.~Strzelecki}
\author[WFAIS]{D.~Trybek}
\author[WFAIS,PAN]{A.~Wieczorek}
\author[WFAIS]{M.~Zieli\'nski}
\author[SWIERK]{W.~Wi\'slicki}
\author[UMCS]{B.~K.~Zgradzi\'nska}
\author[WFAIS]{P.~Moskal}

\address[SWIERK0]{High Energy Physics Division, National Centre for Nuclear Research, 05-400 Otwock-\'Swierk, Poland}
\address[WFAIS]{Faculty of Physics, Astronomy and Applied Computer Science,
 Jagiellonian University, 30-059 Cracow, Poland}
\address[UMCS] {Department of Nuclear Methods, Institute of Physics, Maria Curie-Sklodowska University, 
Pl. M. Curie-Sklodowskiej 1, 20-031 Lublin, Poland}
\address[PAN]{Institute of Metallurgy and Materials Science of Polish Academy of Sciences, Cracow, Poland.}
\address[SWIERK]{\'Swierk Computing Centre, National Centre for Nuclear Research, 05-400 Otwock-\'Swierk, Poland}

\begin{abstract}
The  Jagiellonian-PET (J-PET) collaboration is developing a prototype TOF-PET detector based on long    polymer scintillators. This novel approach exploits the excellent time properties of the plastic scintillators,     which permit very precise time measurements. The very fast, FPGA-based front-end electronics and the    data acquisition system, as well as, low- and high-level  reconstruction algorithms  were specially       developed to be used  with the J-PET scanner. The TOF-PET data processing and reconstruction are time     and resource demanding operations, especially in case of a large acceptance detector, which works in    triggerless data acquisition mode. In this article, we discuss the parallel   computing methods applied to    optimize the data processing for the J-PET detector. We begin with general concepts of parallel computing
    and then we discuss several applications of those techniques in the J-PET data processing
\end{abstract}

\begin{keyword}
\texttt{DAQ} \sep \texttt{computing} \sep \texttt{TOF-PET}
\end{keyword}

\end{frontmatter}


\section{Introduction}
The Jagiellonian-PET (J-PET) collaboration is developing a prototype TOF-PET detector based on plastic scintillators~\cite{Patents,NovelDetectorSystems,NIM14PM,NIM14LR,NIM15PM,wieczorek,moskalA, kowalski}. The detector is a cylinder made of long scintillator strips. Its large acceptance allows for full 3-D image reconstruction. The main advantage of the J-PET solution is its excellent time resolution (see e.g. results in~\cite{NIM14PM}), which makes it suitable not only for medical purposes, but also for precise studies of the discrete symmetries in positronium systems~\cite{kaminska}. The TOF-PET data processing and reconstruction are time- and resource-demanding operations, especially in case of a large acceptance J-PET detector, which works in the so-called trigerless mode, in which all events (digitized time and amplitudes) from the front-end electronics (FEE) are stored to disks without any master trigger condition applied~\cite{korcyl}. Next, the collected raw data undergoes a process of low- and high- level reconstructions. The registered data is first transformed into the hit positions in the scintillator modules, and in the next step the hits are combined to form the Lines of Response(LOR). In the last stage, the image reconstruction procedures are used to obtain the final image based on the set of LORs. In order to efficiently process this high data stream, parallel computing techniques have been applied at several levels  of the data collection and reconstruction.

\section{Parallel processing}
The parallel processing can be defined as a type of computation in which the task is divided into independent subtasks, which are then calculated simultaneously, by several computing resources. The results of the individual computations are merged together. Parallelization techniques can be classified according to several criteria, e.g. instruction-level parallelization corresponds to the simultaneous performance of several operations in the computer program. In the case of the data parallelization, the data set is distributed among many computing nodes, while in case of the task parallelization the code is divided into threads and executed across the computing nodes. Typically, to take advantage of the parallelization, the software procedures must be designed in a special way, e.g. by using dedicated programming environments and libraries such as MPI~\cite{mpi}, OpenMP~\cite{openmp} or CUDA~\cite{cuda}. Overview of different parallelization techniques can be found in~\cite{grama}.
In the past parallel processing was the domain of high-performance computing by means of supercomputers. However, thanks to a very fast development of the overall performance of the CPUs , to keeping the prices relatively low and the introduction of new techniques such as multi-core processors, the parallelization has become more accessible and popular in many different fields. 
Apart from the CPU processing, recently, even more efficient technologies such as Graphical Processor Units (GPU) or Field Programmable Gate Array (FPGA) gained a lot of attention.
In the J-PET project,  parallelization by using multi-core CPUs, GPUs and FPGAs are used at different stages of data processing.

\section{FPGA processing in FEE and Data Acquisition System}
FPGA is a programmable silicon chip which combines two important features: on one hand, the FPGA is reprogrammable, therefore any logic can be implemented and changed if needed in hardware description languages such as Verilog or VHDL. On the other hand, the compiled program is translated to the set of physical connections between the logical arrays, therefore it is really the hardware realization of the designed logic with the functionality of  the real-time speed  processing, analogically to the one offered by the dedicated ASIC processors. Finally FPGA chips are perfect for the parallelization and very cost-effective.
The FPGA devices are the core computing nodes of the JPET FEE and Data Acquisition System (DAQ)~\cite{korcyl}. The J-PET FEE was designed in view of sampling in the voltage domain of  very fast signals at many levels, with a raising time of about 1 ns~\cite{palka}. A novel technique for precise measurement of time and charge is based solely on FPGA  devices and few satellite discrete electronic components. One computing board (called Trigger Readout Board – TRB) consists of five Lattice ECP3-150 FPGAs. Four FPGAs are used as time-to-digital converters and one as a central FPGA  node that steers the whole board. The multiple computing boards are interconnected via network concentrators.  The global time synchronization is provided through a reference channel. The J-PET DAQ system allows for continuous data recording over the whole measurement period. In total, more than 500  channels with 1Gb/s data rates can be read. The overall constant read-out rate is equal to 50 kHz, while reducing the dead time to the level of tens of ns. 
The described triggerless mode of operation allows to store every event without information loss due to preliminary selection. On the other hand, a significant amount of disk storage is needed (about TB per measurement) to save the data, whereas most of the currently registered events contain useless noise information only. In order to reduce the data flow and to eliminate background events a new Central Controller Module (CCM) is introduced as an intermediate computing node between the TRB boards and the disk storage. The CCM is being developed based on Xilinix Zynq chip which contains FPGA integrated with the ARM processor. It is capable of hardware processing up to 16 Gbit ethernet stream in parallel as well as online filtering of the data. Moreover, it is even possible to implement some online reconstruction algorithms. Finally, the online monitoring with a dedicated data substream will be added.

\section{Data parallelization in the low-level reconstruction framework}
The raw data stored on the disks, is processed in the J-PET framework, which serves as a programming environment which provides useful tools  for various reconstruction algorithms, calibration procedures and which standardizes the common operations, e.g: input/output process and more. It also provides the necessary information about run conditions, geometry and electronic setups by communicating with the parameter database. The architecture of the analysis framework was already described in~\cite{krzemien1,krzemien2}. In this paragraph we will describe the important parts in the context of understanding framework parallelization. In the J-PET framework, the analysis chains are decomposed into series of standardized modular blocks. Each module corresponds to a particular computing task, e.g. reconstruction algorithm or calibration procedure, with defined input and output methods. The processing chain is built by registering chosen modules in the JPetManager, which is responsible for the synchronization of the data flow between the modules. The framework parallelization is implemented by using the PROOF~\cite{proof} (Parallel ROOT Facility) extension for the ROOT library~\cite{root}. PROOF enables parallel file processing on cluster of computers or many core machines. In the case of the J-PET framework the multi-core processing was tested. Two options are being developed. The first solution is a realization of data parallel computing. First, a set of chosen computing tasks ,in the form of processing chain, is registered in the JPetManager as described before. The same processing chain will be multiplied and executed in parallel for every input file provided. This approach assumes that the input files can be analyzed independently. In the second mode, a single processing chain can contain modules (subtasks) that can operate in parallel. This solution is currently being implemented.

\section{Parallelization at the image reconstruction level}
The final output of the low-level reconstruction phase is a reconstructed set of LORs that is provided as the input data for the image reconstruction procedures. The most popular  approach based on iterative algorithms derived from Maximum Likelihood Estimation Method  (MLEM)~\cite{shepp} has been adopted. The available time-of-flight information is incorporated to improve the accuracy  and the quality of the reconstruction. In order to reduce the processing time, parallelization techniques are applied. Currently two implementations are used. The first solution exploits the processing capability of Graphical Processing Units (GPU). The efficient image reconstruction using list-mode MLEM algorithm with approximation kernels was implemented for GPU~\cite{bialas}. Here, the CUDA platform was adopted. 
The second approach is a full 3-D reconstruction based on a multi-core CPU architecture~\cite{slomski}. In this  case, the most time-consuming operations such as projection and back-projections are parallelized. The code is based on the OpenMP library. For the current test implementation, the time of one MLEM iteration, processed on 40 cores with 128 GB, is about 70 minutes, when using the large field-of-view (88 cm x 88cm x 50 cm) with a binning of 0.5 cm and 1 degree. Typically about 10 iterations are enough to reach MLEM optimal reconstruction point.
\section{Summary and Outlook}
In order to reduce the processing time of the data flow, we use the parallel computing approach on several stages. We presented the implemented solution in the FFE and DAQ level based on the FPGA chips. Also, the multi-core CPU-based and GPU-based algorithms are used for the low-level and high-level reconstructions. Currently, works are ongoing to further reduce the processing time, e.g. by implementing the online event filters. Apart from the presented computing schemes, in which the data processing is performed locally, several remote processing concepts are considered as a replacement to the traditional in-site computing. The basic idea is to carry outs the resource-heavy computations remotely by using cloud or grid-computing~\cite{wislicki}.
\section{Acknowledgements}
We acknowledge technical and administrative support by T. Gucwa-Ry\'s, A. Heczko,
M. Kajetanowicz, G. Konopka-Cupia\'l, W. Migda\'l, and the financial support by the Polish National Center for Development and Research through grant No. INNOTECH-K1/IN1/64/159174/NCBR/12, the Foundation for Polish Science through MPD programme and the EU, MSHE Grant No. POIG.02.03.00-161 00-013/09 and Doctus – the Malopolska PhD Scholarship Fund.


%

\end{document}